# NONEQUILIBRIUM THERMODYNAMICS AND LIFETIME OF PHYSICAL SYSTEMS

V.V.Ryazanov*, S.G.Shpyrko

*Corresponding author. E-mail address: vryazan@kinr.kiev.ua

Institute for Nuclear Research, pr. Nauki, 47, 03068, Kiev, Ukraine

To describe the nonequilibrium states of a system we introduce a new thermodynamic parameter - the lifetime of a system. The statistical distributions which can be obtained out of the mesoscopic description characterizing the behaviour of a system by specifying the stochastic processes are written down. The expressions for the nonequilibrium entropy, temperature and entropy production are obtained, which at small values of fluxes coincide with those derived within the frame of extended irreversible thermodynamics. The expressions generalizing the Maxwell-Cattaneo relations of extended irreversible thermodynamics, and their analogues for mass transfer and chemical reactions are obtained.

## I. INTRODUCTION

In papers [1-3] the linear nonequilibrium thermodynamics (*LIT*) was developed which found its application in the variety of physical problems. But the linear irreversible thermodynamics possesses a number of restrictions which present certain obstacles for adequately describing such phenomena as the propagation and absorption of ultrasound in liquids, density profile of shock waves in gases etc. The attempts to overcome these difficulties led to creating the extended irreversible thermodynamics (*EIT*) [4]. If in the frame of *LIT* [1-3] the local entropy density is merely a function of conserving quantities such as local energy and mass densities, in *EIT* a set of additional variables is added thereto which are chosen as densities of fluxes in a system (nonconserving dissipative values). Phenomenological laws appear to be  nonstationary equations of motions, the entropy flux - containing nonlinear terms, the velocity of propagation of perturbations is finite in the contrast to the infinite velocity (e.g. of propagating heat waves) predicted by classical thermodynamics. The Fourier law $\vec{q}=-\lambda\vec{\nabla} T$ (where $\vec{q}$ is the heat flux, $\lambda$ is the heat conductivity coefficient, $T$ is absolute temperature) of the classical *LIT* is replaced by the Maxwell-Cattaneo equation

$$\vec{q}=-\lambda\vec{\nabla} T-\tau_q\partial\vec{q}/\partial t, \qquad (1)$$

where $\tau_q$ is the time of the flux correlation.

Characterizing the nonequilibrium state by means of an additional parameter related to the deviation of a system from the equilibrium (field of gravity, electric field for dielectrics etc) was used in [5]. In the present paper it is suggested the new choice of such an additional parameter as the lifetime of a physical system which is defined as a first-passage time till the random process *y(t)* describing the behaviour of the macroscopic parameter of a system (energy, for example) reaches its zero value. The lifetime is thus a random process which is slave (in terms of the definitions of the theory of random processes [6]) with respect to the master process *y(t)*,

$$\Gamma_x=\inf\{t: y(t)=0\},\ y(0)=x>0. \qquad (2)$$

The characteristics of $\Gamma$ depend on those of *y(t)*. The solution to the lifetime problem stands close to the problem of Kramers [7] (overcoming the potential barrier), althougs the formulation is more general.

In work [8] lifetimes of the system are considered as the random moments of a first-passage time till the random process describing system certain border, for example, zero value reaches. In [8] are received approached exponential expressions for the probability density function (with single parameter) and probability distribution of lifetime, the accuracy of these expressions is estimated. In works [9,10] the lifetimes of molecules are investigated, the affinity of real distribution for lifetime and approached exponential model is shown.

In [11] the auxiliary weight function $p_q(y)=\varepsilon exp\{-\varepsilon y\}$ from Zubarev Nonequilibrium Statistical Operator (*NSO*) [12] is understood as the distribution function of a system lifetime from the time of its birth $t_0$ till the current moment *t*. It was marked in [13] that Zubarev form of *NSO* incorporates the irreversibility from the outset using an ad hoc non-mechanical hypothesis. In the interpretation of [11] these hypotheses are related to the existence lifetime of a system. This is the reason why one can use the lifetime as a random thermodynamical parameter. The logarithm of Zubarev *NSO* [12] is

$$ln\rho(t)=\int_0^\infty \varepsilon exp\{-\varepsilon y\}ln\rho_q(t-y,-y)dy=ln\rho_q(t,0)+\int_0^\infty exp\{-\varepsilon y\}(dln\rho_q(t-y,-y)/dy)dy,$$

where $\rho_q$ is the quasiequilibrium distribution [12]; in the interpretation of [11] $\varepsilon=1/<\Gamma>$, $<\Gamma>=<t-t_0>$. If the value $(dln\rho_q(t-y,-y)/dy)$ has weak *y*-dependence then

$$\int_0^\infty exp\{-\varepsilon y\}(dln\rho_q(t-y,-y)/dy)dy \approx <\Gamma>(dln\rho_q(t-y_1,-y_1)/dy_1),$$

which corresponds to the expression (7), but instead of $<\Gamma>(dln\rho_q(t-y_1,-y_1)/dt)$ there stand the random value $\Gamma$ and Lagrange multiplier $\gamma$ in: $\gamma\Gamma+lnZ(\beta,\gamma)/Z(\beta)$.

The lifetime seems to be a value of more general character than the fluxes are (this fact being the prerequisite of formulating a theory which would include the *EIT* as its partial case). The history of a system is the succession of repeating "busy periods" - lifetimes in the sense of the definition (2), and idle periods (like it is the case of the queues systems) which are marked by the observable events of lifetime finishing and system regenerating. The internal time in the system is related to the existence of the nonzero number of elements within. The principal consideration grounding the introduction of the lifetime as a physical parameter is an empirical fact of all systems having limited time of existence.

In the present paper we provide the statistical foundations for the thermodynamic relations and introduce the distribution for the lifetime of a system (Section 2). The main assumptions, namely, concerning the explicit lifetime distribution and the relation between its thermodynamic conjugate and the fluxes are given in the Section 3,4. The explicit expressions for the temperature, entropy and entropy production are written down and the stability analysis is performed in Section 4. In the Conclusion various aspects of proposed method for describing complex systems are treated.

## II. LIFETIME DISTRIBUTIONS

Lets consider a macroscopic observable $E$ (it might be something different from energy as well) being distributed with function $P(E)=p_E(x)$. Using the maximum-entropy inference [14] one can reconstruct out of the macroscopic distribution $P(E)=p_E(x)$ the microscopic probability density in the phase space z. The constructing of Gibbs microcanonic distribution corresponds to imposing the additional condition of the equiprobable distribution of all possible microstates. This "maxent" (i.e. related to the maximum-entropy principle) procedure corresponds [15] to the process of coarsening and means physically the loss of information about further details of the behaviour of a system. Lets suppose that bringing a system in the nonequilibrium state violates the equiprobable distribution of equilibrium. We thus introduce a novel parameter $\Gamma(z)$ supposing its observability. We can introduce as well the cells of the extended phase space with constant values of the set $(E,\Gamma)$ within (instead of the cells with constant values of $E$). The structure factor $\omega(E)$ is thus replaced by $\omega(E,\Gamma)$ - the volume of the hyperspace containing given values of $E$ and $\Gamma$. If $\mu(E,\Gamma)$ is the number of states in the phase space which have the values of $E$ and $\Gamma$ less than given numbers, then $\omega(E,\Gamma)=d^2\mu(E,\Gamma)/dEd\Gamma$. It is evident that $\int\omega(E,\Gamma=y)dy=\omega(E)$. The number of phase points between $E,E+dE$; $\Gamma,\Gamma+d\Gamma$ equals $\omega(E,\Gamma)dEd\Gamma$. We make use now of the principle of equiprobability applied to the extended cells $(E,\Gamma)$.

The standard procedure (e.g. [15], [16]) allows one to write down the relation between the distribution density $P(E,\Gamma)=p_{E\Gamma}(x,y)$ and microscopic (coarse-grained) density $\rho(z;E,\Gamma)$

$$P(E,\Gamma)=p_{E\Gamma}(x,y)=\int\delta(E-E(z))\delta(\Gamma-\Gamma(z))\rho(z;E,\Gamma)dz, \qquad (3)$$

where

$$P(E,\Gamma)=p_{E\Gamma}(x,y)=\rho(z;E,\Gamma)\omega(E,\Gamma)|_{E=x,\Gamma=y}. \qquad (4)$$

Lets introduce the conditional distribution density

$$P(\Gamma|E)=p_\Gamma(y|x)=p_{E\Gamma}(x,y)/p_E(x)=P(E,\Gamma)/P(E), \qquad (5)$$

where

$$P(E)=p_E(x)=\int_0^\infty \rho(z;E,\Gamma=y)\omega(E,y)dy . \qquad (6)$$

The value $E$ should be understood in a wider context than merely the energy. In the general case as $P(E)$ one could take an analog to the quasiequilibrium distribution [12]. Now we make an assumption about the form of $\rho(z;E,\Gamma)$ :

$$\rho(z;E=x,\Gamma=y)=exp\{-\beta x-\gamma y\}/Z(\beta,\gamma) , \qquad (7)$$

where

$$Z(\beta, \gamma)=\int exp\{-\beta x-\gamma y\}dz=\iint dxdy\,\omega(x,y)exp\{-\beta x-\gamma y\} \qquad (8)$$

is the partition function, $\beta$ and $\gamma$ are Lagrange multipliers satisfying the equations for the averages

$$<E>=-\partial \ln Z/\partial \beta |_\gamma; \qquad <\Gamma>=-\partial \ln Z/\partial \gamma |_\beta. \qquad (9)$$

We did not concretize yet the parameter $\Gamma(z)$. Now let it be the lifetime of a system (2). Introducing $\Gamma$ means effective account for more information than merely in linear terms of the canonical distribution.

The linearity of the exponent in (7) allows one to relate the expression for the partition function $Z(\beta, \gamma)$ with the Lagrange transforms of the macroscopic distributions $P(x)$, $P(\Gamma|x)$, $P(\Gamma, x)$

$$E(e^{-\theta x})\equiv Q(e^{-\theta})=\int e^{-\theta x}P(x)dx ; \qquad (10)$$

$$E(e^{-s\Gamma(x)}) \equiv L(x,s) = \int e^{-sy} P(\Gamma=y|x) dy ; \qquad (11)$$

$$G(\theta, s) \equiv \int e^{-\theta x - s y} P(E=x, \Gamma=y) dx dy = L(-\partial/\partial\theta, s) Q(e^{-\theta}) , \qquad (12)$$

where $E(...)$ means averaging. The expressions (10)-(13) can be easily obtained explicitly using the known model of the process $y(t)$ (see Appendix 1).

From the expressions (4), (12) and (7-8) it is easy to derive the relation between microdynamics (containing in the partition function $Z(\beta, \gamma)$) and the Laplace transforms of the macrovariable and its lifetime

$$G(\theta, s) \equiv Z(\beta+\theta, s+\gamma)/Z(\beta, \gamma) . \qquad (13)$$

Let us choose some reference point $(\beta_0, \gamma_0)$ in the space of the Lagrange parameters. If the macroscopic model holds for this point one can get (using (13)) the value of the nonequilibrium partition function for the other point $(\beta, \gamma)$,

$$Z(\beta, \gamma) = Z(\beta_0, \gamma_0) G(\beta-\beta_0, \gamma-\gamma_0) . \qquad (14)$$

The value $Z(\beta_0, \gamma_0)$ is merely a number characterizing the "bound part" of the entropy of the microscopic motion. The multiplier $G$ corresponds to the free information derivable when concretizing the stochastic process [17].

Let us underline the principal features of the suggested approach.

1. We introduce a novel variable $\Gamma$ which can be used to derive additional information about a system in the stationary nonequilibrium state. We suppose that $\Gamma$ is a measurable quantity at macroscopic level, thus values like entropy which are related to the order parameter (principal macroscopic variable) can be defined. At the mesoscopic level the variable $\Gamma$ is introduced as a variable with operational characteristics of a random process slave with respect to the process describing the order parameter.

2. We suppose that thermodynamic forces $\gamma$ related to the novel variable can be defined. One can introduce the "equations of state" $\beta(<E>,<\Gamma>)$, $\gamma(<E>,<\Gamma>)$.

Thus we introduce the mapping (at least approximate) of the external restrictions on the point in the plane $\beta, \gamma$.

3. We suppose that a "refined" structure factor $\omega(E,\Gamma)$ can be introduced which satisfies the condition $\int\omega(E,\Gamma=y)dy=\omega(E)$ (ordinary structure factor). This function (like $\omega(E)$) is the internal (inherent) property of a system. At the mesoscopic level we can ascribe to this function some inherent to the system (at given restrictions $(\beta_0,\gamma_0)$) random process. The structure factor has the meaning of the joint probability density for the values $E,\Gamma$ understood as the stationary distribution of this process. Provided the "reper" random process for the point $(\beta_0,\gamma_0)$, one can derive therefrom the shape of the structure function. If we model the dependence of the system potential of the order parameter by some potential well, the lifetime distribution within one busy period and probabilities $\omega(E,\Gamma)$ can be viewed as distributions of the transition times between the subset of the phase space (possibly of the fractal character) corresponding to the potential well, and the subset corresponding to the domain between the "zero" and the "hill" of the potential where from the system will roll down to the zero state. To determine the explicit form of $\Gamma$ (at $(\beta_0,\gamma_0)$) the algorythm of the asymptotic phase coarsening of complex system is used (Section 3).

4. It is supposed that at least for certain classes of influences the resulting distribution has the form (4), (7), that is the change of the principal random process belongs to some class of the invariance leading to this distribution which explains how one can pass from the process in the reper point $(\beta_0,\gamma_0)$ (for example, in equilibrium when $\gamma=0$ and $\beta=1/k_BT$) to a system in an arbitrary nonequilibrium stationary. The thermodynamic forces should be chosen so that the distribution lead to new (measurable) values of $(<E>,<\Gamma>)$.

## III. GENERALIZED LIFETIME THERMODYNAMICS

If $\gamma=0$ and $\beta=\beta_0=(k_BT_{eq})^{-1}$, where $k_B$ is the Boltzmann constant, $T_{eq}$ is the equilibrium temperature, then the expressions (7-8) yield the equilibrium Gibbs distribution. One can thus consider (7-8) as a generalization of the Gibbs statistics to cover the nonequilibrium situation. Such physical phenomena as the metastability, phase transitions, stationary nonequilibrium states are known to violate the equiprobability of the phase space points. The value $\gamma$ can be regarded as a measure of the deviation from the equiprobability hypothesis. In general one might choose the value $\Gamma$ as a subprocess of some other kind as chosen above.

Let us suppose that the process $y(t)=E$ is the energy of a system (equivalently one could choose the particle number, pulse etc). Further detalization will require the concretization of the $\Gamma$ distribution and the interpretation of the Lagrange parameters $\beta$ and $\gamma$. The Lagrange parameter $\beta$ is supposed to be (likewise the equilibrium Gibbs statistics)

$$\beta = 1/k_B T ,  \qquad (15)$$

where $T$ is the average (over the body volume) local equilibrium temperature. Since at fluxes $\vec{q} \neq 0$ the temperature is not the same all over the bulk of the body, one can define $T$ in a system with volume $V$ as the volume average, i.e. $T = V^{-1} \int_V T(r,t) dr$ (the same definition was used in [18]). Of cource, the thermodynamic description itself is supposed to be already coarse-grained. To get the explicit form of the $\Gamma$ distribution we shall use the general results of the mathematical theory of phase coarsening of the complex systems [19] (Appendix 2), which imply the following distribution of the lifetime for coarsened random process (see also [8]):

$$p_\Gamma(y) = \Gamma_0^{-1} exp\{-y/\Gamma_0\} . \qquad (16)$$

The values $\Gamma_0$ are averaging of the residence times and the degeneracy probabilities over stationary ergodic distributions (in our case - Gibbs distributions). The physical reason for the realization of the distribution in the form (16) is the existence of the weak ergodicity in a system. Mixing the system states at big times will lead to the distributions (16).

As we note in Section 2, the structure factor $\omega(E,\Gamma)$ has a meaning of the joint probability density of values $E,\Gamma$. For the distribution (16) the function $\omega(E,\Gamma)$ from (4), (6), (8) take on the form:

$$\omega(E,\Gamma=y)=\omega(E)\Gamma_0^{-1}exp\{-y/\Gamma_0\} \ . \tag{17}$$

Substituting into the partition function (8) yield

$$Z(\beta,\gamma)=Z(\beta)(1+\gamma\Gamma_0)^{-1} \ , \tag{18}$$

where $Z(\beta)=\int\omega(E=x)exp\{-\beta x\}dx$ is the Gibbs partition function.

We have from (9) and (18) when $\Gamma_\gamma=-\partial\ lnZ(\beta,\gamma)/\partial\gamma;\ \ \Gamma_0(V)=\Gamma_\gamma(V)|_{\gamma=0}$;

$$\Gamma_\gamma=\Gamma_0/(1+\gamma\Gamma_0), \quad \gamma=1/\Gamma_\gamma-1/\Gamma_0 \ , \tag{19}$$

that is $\gamma$ is the difference between the inverse lifetimes of the open system $1/\Gamma_\gamma$ and the system without external influences $1/\Gamma_0$ which can degenerate only because of its internal fluctuations. The value $\gamma$ is thus responsible for describing the interaction with the environment and its existence is the consequence of the open character of a system. When defining $\gamma$ one should take into account all factors which contribute to the interaction between the system and the environment. If one denotes in (19) $x=\gamma\Gamma_0$, then $x=x_1+x_2+...+x_n$, where the value $x_i$ is determined by the flux labelled by the index $i$.

Lets introduce the nonequilibrium entropy corresponding to the distribution (7) in the analogy to [12] by the relation

$$S/k_B=-<ln\rho(z;E,\Gamma)>=\beta<E>+\gamma<\Gamma>+lnZ(\beta,\gamma) \ . \tag{20}$$

For the case of several potential wells with the same minima positions (which means the case of a system with several ergodic states) one can expect the appearance of the Erlang-type distributions instead of the exponential one (16) [19]. More precise approximations can also be used.

### IV. ONE-PHASE SYSTEMS (SYSTEMS POSESSING ONE CLASS OF THE STABLE STATES); HEAT CONDUCTIVITY, MASS TRANSFER AND CHEMICAL REACTIONS

Substituting into (20) the expressions (18) and (19), we have

$$S/k_B = S_\beta/k_B + x/(1+x) - \ln(1+x); \quad S_\beta/k_B = \beta \langle E \rangle + \ln Z_\beta; \quad x = \gamma \Gamma_0. \tag{21}$$

From (20) treating $E$ and $\Gamma$ as variables:

$$dS = k_B \beta \, dE + k_B \gamma \, d\Gamma. \tag{22}$$

From (19) $d\Gamma = -\Gamma^2 d\gamma + (\Gamma/\Gamma_0)^2 d\Gamma_0$. Substituting into (22) we find from the Maxwell relations that $\partial \Gamma_0/\partial E$, $\partial \Gamma_0/\partial R$ are proportional to $\partial \Gamma_0/\partial \vec{q} = 0$ (since $\Gamma_0$ by definition corresponds to the state with $\vec{q} = 0$ and $\Gamma_0$ independent on $\vec{q}$). Thats why $d\Gamma_0 = 0$, and

$$dS = k_B \beta \, dE - k_B x \, dx/(1+x)^2, \tag{23}$$

which coincides with the differential of (21).

Now we determine the explicit form for $\gamma$ and $x$. For this purpose we will recall the expressions of the *EIT* [4]:

$$S = S_\beta - \int_V \rho \alpha_q (\vec{q}\vec{q}) d\vec{r}/2; \quad \alpha_q = \tau_q/\rho \lambda T^2 \tag{24}$$

($\vec{q}$ and $\tau_q$ are values from (1)). Comparing with (21) at small $\vec{q}$: $S = S_\beta - k_B x^2/2$; $k_B x_q^2 = \int_V \rho \alpha_q (\vec{q} \; \vec{q}) d\vec{r}$;

$$x = \pm \left( \int_V \rho \alpha_q (\vec{q} \; \vec{q}) d\vec{r}/k_B \right)^{1/2} = \gamma \Gamma_0. \tag{25}$$

The derivative $\partial x/\partial \vec{q}|_{E,R}$ is written as $\int_V [\delta x/\delta \vec{q}(\vec{r})|_{E,R}]d\vec{r} = \int_V \rho\, \alpha_q \vec{q}\, d\vec{r}/k_B x$, where $R$ is the linear dimension of the system. The expressions for $\partial x/\partial E|_{q,R}$, $\partial x/\partial R|_{E,q}$ can be obtained in the same fashion. Then from (23)

$$dS = \{k_B\beta - \int_V [\partial(\rho\,\alpha_q)/\partial u](\vec{q}\,\vec{q})d\vec{r}/\rho V2(1+x)^2\}dE - \int_V \rho\,\alpha_q \vec{q}\, d\vec{r}\, d\vec{q}/(1+x)^2 - \quad (26)$$
$$2x^2 k_B dR/R(1+x)^2.$$

The sign "-" in (25) is chosen if the incoming flux is directed towards the system (the system is being heated). Outcoming flux $\vec{q}_-$ and incoming flux $\vec{q}_+$ differ in their times: we consider $\vec{q}_+$ to be the first flux. In (25) we suppose the finite velocity of the heat propagation in the system. Thus the linear dimensions enter to the characteristics of the "point" system. For continuous systems $R=L$, where $L$ is the size of the "point" of the continuous medium in the continuous description [16, 18]. In the frame of the kinetic theory $L=l_{ph}=l_1(\varepsilon_1)^{1/2} \ll l_1$, $\varepsilon_1 = nr_0^3$, where $n$ is the density, $r_0$ is the diameter of the atom, $l_1$ is the free path length. In the gasodynamic description $L=l_{ph}^G \sim L^G/N^{1/5}$, where $L^G$ is the size of the system, $N$ is the particle number. One should take into account the source density $\sigma$ when describing the reaction systems. For example, in the nuclear reactors one must take into account the terms responsible for the fuel sources. But even in the equations for the internal energy one should account for the sources if external fields and/or inner heterogeneities are present [3]. Let the system be open with respect to the quantity $A=\int_V \rho\, a dV$, for which the balance equation $\partial(\rho\, a)/\partial t + \vec{\nabla}\vec{J}^0_a = \sigma_a;\quad \vec{J}^0_a = \rho_a \vec{v}_a$ [3] holds ($\vec{v}$ is the velocity, $\rho = M/V$ being the mass density); $div\,\vec{J}^0_a \approx (J^0_{a-} + J^0_{a+})/R$, where $R$ is the "size" of the point in the continuous description [18]. Then the value $J^0_{a+} - R\sigma_a = -J^0_{a-} - R\partial(\rho\, a)/\partial t$ is proportional to $\gamma$ from (19). And the value $y_a = (J^0_{a+} - R\sigma_a)/Ra$, multyplied by the time parameter $t_{0a}$ (from (27)), coincides with $x = \gamma\, \Gamma_0$. The equation for the specific energy $u$ has the form $\rho\, du/dt + \vec{\nabla}\,\vec{J}_q = 0;\ E = \int_V \rho\, u dV;\ \vec{J}_q = \vec{J}_u = \vec{q}$ is the heat flux. We consider the case of the constant mass density.

If the fluxes $q$ depend only weakly on the spatial variables then (25) is cast as

$$x_q \approx \pm(\int_V \rho\, \alpha_q\, d\vec{r}\, /k_B)^{1/2}(\vec{q}\, \vec{q}\,)^{1/2} \approx t_{0q}S_a q/E = t_{0q}q/\rho\, uR = t_{0q}y_q\,; \qquad (27)$$

$$t_{0q} \approx E(\int_V \rho\, \alpha_q\, d\vec{r}\, /k_B)^{1/2}/S_a;\quad y_q = q/\rho\, uR;\quad E \approx \rho\, uV,$$

where $S_a \sim R^2$ is the surface of the system, $V \sim R^3$ is its volume, $E$ is the energy value at the homeokinetic plateau; $q=q_+=(\vec{q}_+,\vec{n})$ ; $q_+$ is the scalar product of the incoming flux vector $\vec{q}_+$ and the unity length normal vector directed outwards to the system surface $\vec{n}$. One can show that for the model with the homeokinetic plateau [20] the time $t_0=t_{0q}$ has its physical meaning of the degeneracy time of a system, i.e. the time during which a system will pass from the homeokinetic plateau to the degenerate state with zero energy.

From (23) we have

$$dS = k_B[\beta + x^2/E(1+x)^2]dE - [k_B x^2/q(1+x)^2]dq - [k_B 2x^2/R(1+x)^2]dR\,. \qquad (28)$$

Consider now the stability of the thermodynamic system. To ensure its stability the condition $\delta^2 S \leq 0$ should satisfy. We have:

$k_B^{-1}\partial^2 S/\partial E^2|_{q,R} = \partial \beta/\partial E - x^2(3+x)/E^2(1+x)^3$; $k_B^{-1}\partial^2 S/\partial q^2|_{E,R} = x^2(x-1)/q^2(1+x)^3$; $k_B^{-1}\partial^2 S/\partial R^2|_{E,q} = 2x^2(x-3)/R^2(1+x)^3$; $k_B^{-1}\partial^2 S/\partial E\partial q = 2x^2/qE(1+x)^3$; $k_B^{-1}\partial^2 S/\partial q\partial R = -4x^2/qR(1+x)^3$; $k_B^{-1}\partial^2 S/\partial E\partial R = 4x^2/ER(1+x)^3$; $\partial \beta/\partial E = -1/k_B T^2 c$; $c = \partial E/\partial T$ is the heat capacity. The nonequilibrium heat capacity $c_N = \partial E/\partial \theta$ is likewise defined. If $R=const$ (variables are $E$ and $q$) the condition $\delta^2 S \leq 0$ holds for $\partial^2 S/\partial E^2|_{q,R} \leq 0$, $\partial^2 S/\partial q^2|_{E,R} \leq 0$. These conditions satisfy when $-1/T^2 c - k_B x^2(3+x)/E^2(1+x)^2 \leq 0$; $k_B x^2(x-1)/q^2(1+x)^3 \leq 0$. One more condition satisfies if $\partial^2 S/\partial E^2|_{q,R}\, \partial^2 S/\partial q^2|_{E,R} - (\partial^2 S/\partial E\partial q)^2 \geq 0$. Hence

$$|x| \leq (1 + k_B T^2 c/E^2)^{-1/2}\,. \qquad (29)$$

It is seen from (29) that at $q<0$, $x<0$, $c>0$, $|x|<1$ that is the denominator in (19), (28) does not equal to zero in the domain of the thermodynamic stability and the expression for $\Gamma$ converge. The same relations are written for $E=const$ in the variables $q,R$ and for $q=const$ in terms of $E,R$.

Determining in (28) $dS/dt$ we find the entropy balance equation $(dS/dt)/V = -\vec{\nabla}\vec{j}_S + \sigma_S$, $S = \int_V \rho s d\vec{r}$, where for $\rho = const$, $\vec{j}_S = \theta^{-1}\vec{q}$;

$$\sigma_S = \vec{q}\,\vec{\nabla}\,\theta^{-1} - k_B x^2 (dq/dt)/qV(1+x)^2 + k_B(dR/dt)[3u\beta\rho + x^2 V^{-1}(1+x)^{-2}]/R.$$

Comparing entropy production $\sigma_S$ and $\sigma_S = \vec{q}\,\vec{q}/\lambda T^2$ [3], we shall find

$$\lambda T^2 k_B x^2 (d\vec{q}/dt)q^{-2}V^{-1}(1+x)^{-2} + \vec{q} = \lambda T^2 \nabla \theta^{-1} + \qquad (30)$$
$$(dR/dt)\lambda T^2 k_B[3u\beta\rho + x^2 V^{-1}(1+x)^{-2}]/\vec{q}\,R.$$

The inverse nonequilibrium temperature
$$\theta^{-1} = \partial S/\partial E|_{q,R} = k_B[\beta + x^2/E(1+x)^2] = 1/T + q^2\alpha_q/u[1 + q(\rho\alpha_q V/k_B)^{1/2}]^2$$
coincides with the value $1/\theta$ of the extended thermodynamics [4] at small $q$ (when $(1+x)^{-2} \approx 1$). The value entropy production $\sigma_S$ and the expression (30) when substituting therein (27) take on the form

$$\sigma_S = \vec{q}\,\vec{\nabla}\,\theta^{-1} - \rho q\alpha_q(dq/dt)/(1+x)^2 + (dR/dt)\rho[3uT^{-1} + q^2\alpha_q(1+x)^{-2}]/R;$$
$$(d\vec{q}/dt)\tau_q(1+x)^{-2} + \vec{q} = \lambda T^2 \vec{\nabla}\,\theta^{-1} + (dR/dt)[3u\lambda T\rho/\vec{q} + \vec{q}\,\tau_q(1+x)^{-2}]/R,$$

and coincide with the corresponding expressions of *EIT* and Maxwell-Cattaneo equation (1) at $dR/dt=0$ and small $q$. The same can be derived from (26).

Above we considered the lifetime as a quantity related to the heat in a system. Since the heat transfer in a body is accompanied by the processes of deformation of a continuous medium, the energy dissipation is conditioned not only by the heat transfer, but by the internal friction of a system which is represented by the dissipative part of the stress tensor; thus the full expression for $x$ must have (similar to [21]) the form

$$x = \pm \left(\int_V \frac{d\vec{r}}{k_B}\rho\left[\frac{\tau_q}{\rho\lambda T^2}(\vec{q}\vec{q}) + (\sigma^v:\sigma^v)\frac{\tau_v}{\rho\mu T}\right]\right)^{1/2},$$

where $\sigma^v$ is the viscous stress tensor, $\tau_v$ is the correlation time of viscous stresses, $\mu$ s the shear viscosity. Correspondingly the expressions for $\vec{j}_S$, $\sigma_S$,… acquire more cumbersome form. Similar expressions can be written down for the density changes (if one considers the lifetime for the full mass), for velocity, and for other factors contributing to the expression for the source $\sigma_u$ from the equation for the specific energy density [3] etc.

In the examples considered above one took for $\beta E$ the values $\int dr \beta(r,t) \hat{u}(r)$, where $\hat{u}(r)$ is the dynamical variable of the energy density, $\beta(r,t)=1/k_B T(r,t)$; if we consider the processes with variable mass, we should take as $\beta E$ the quantities $\int dr \beta(r,t) [\hat{u}(r) - \sum_k \mu_k(r,t) \hat{n}_k(r)]$, where $\hat{n}_k(r)$ is dynamical variable of the particle density for the particles of $k$-th kind, $\mu_k(r,t)$ is the chemical potential of the $k$-th particles. Taking into account the chemical reactions like $\sum_{j=1}^{r} \nu_j^+ X_j \xleftrightarrow{k_+} \sum_{j=1}^{r} \nu_j^- X_j$, we get

$$x_\rho = \pm \sum_{k=1}^{r} (\int_V \frac{d\vec{r}}{k_B} [\frac{\tau_{k\,diff}}{L_k V} (\vec{j}_k \vec{j}_k) + (\omega_k \omega_k) \frac{\tau_{kchem}}{L_{kchem} V T}])^{1/2} ,$$

where $x_k = t_{0k}(J_k - \omega_k R)/Rc_k$; $t_{0k} = t_{0kdiff} = c_k R (\tau_{\rho diff}/k_B L_k)^{1/2}$, where $\tau_{\rho diff}$ is the correlation time for the fluxes $\vec{J}_k$; $J_k$ is the normal projection of $\vec{J}_k$ (the incoming flux into the system of the size $R$) to the surface; $L_k = D_k T/(\partial \mu_k /\partial c_k)_{T,P}$; $\mu_k$ is the chemical potential ($=\mu_{0k}$ without perturbation); $L_k$ is Onsager coefficient from $\vec{J}_i = -L_i \vec{\nabla}(\mu_i/T)_{T,P}$. We consider a multicomponent mixture with mass densities of the components $\rho_k$ and concentrations $c_k = \rho_k/\rho$; $k=1,...,r$; $\sum_{k=1}^{r} \rho_k = \rho$; $c_n = N_n/V$, where $r$ is the total number of particle types involved into reactions, $N_n$ is the number of particles of the $n$-th type. The balance equation for $c_j$ has the form $\partial c_j/\partial t = -\vec{\nabla} \vec{J}_j + \omega_j$, where $\vec{J}_j = -D_j \vec{\nabla} c_j$ is the diffusive flux of the $j$-th component, $D_j$ – its diffusion coefficient, $\omega_j = k_+ \prod_{n=1}^{r} c_n^{\nu_n^+} - k_- \prod_{n=1}^{r} c_n^{\nu_n^-}$ the rate of the $j$-th reaction per unit volume; $k_+$ and $k_-$ are reaction constants; $\nu_n = \nu_n^- - \nu_n^+$ are stechiometric coefficients [3, 22]. Besides the value $t_{0kdiff}$ one can also introduce the time $t_{0kchem} = (\alpha_\rho$

$_{kchem}V/k)^{1/2}$, $\alpha_{\rho\ kchem}=\tau_{kchem}/(VTL_{kchev})$; $L_{kchem}=c_k\omega_+^{eq}/kT$, $\omega_+^{eq}=k_+\prod_{j=1}^{r}(c_j^{eq})^{v^+_j}$ (such choice of kinetic coefficients for chemical reactions where $\omega=LA$ ($A=-\sum_{n=1}^{r}v_n\mu_n=-kTln\prod_{j=1}^{r}(c_j/c_j^{eq})^{vj}$ is chemical affinity) is performed, e.g., in [22,23]), $\tau_{k\ chem}$ is the correlation time of the fluxex $\omega_k$ caused by chemical reactions. Since $t_{0kchem}$ and $t_{0kdiff}$ enter the same balance equation for $c_k$, $t_{0kdiff}=t_{0kchem}$. From this we get the relation between $\tau_{kchem}$ and $\tau_{kdiff}$ : $\tau_{kchem}/(TL_{kchem})=R^2c^2_k\tau_{\rho diff}/L_k$. As thermodynamic quantities lets take $c_k$, $J_k$, $m_k$; $m_k=\sum_{j=1}^{r}\gamma_{kj}\omega_j$; $\omega_j=(\partial_j\rho_k/\partial t)/\gamma_{kj}$ in the $j$-th chemical reaction [3]; $\sum_{k=1}^{K}m_k=0$ (the stechiometric coefficients $v^+_r$ in [22] are cast in another fashion than $v_{kj}$ from [3], [23]). Since $\partial c_k/\partial\rho_k=(1-c_k)/\rho$, $\partial x_k/\partial c_k |_{J_k,m_k}=-x_k(\partial\rho_k/\partial c_k)/\rho_k=x_k/c_k(c_k-1)$; $\partial x_k/\partial J_k|_{c_k,m_k}=x_km_kR/J_k(J_k-R\ m_k)$; $\partial x_k/\partial m_k|_{c_k,J_k}=-Rx_k/(J_k-Rm_k)$. Substituting these expressions into (3), we get

$$(k_B)^{-1}dS=(k_B)^{-1}dS_\beta-(\gamma\Gamma)^2\sum_{k=1}^{r}\gamma_k[dc_k/c_k(c_k-1)+Rm_kdJ_k/J_k(J_k-Rm_k)-Rdm_k/(J_k-Rm_k)]/\gamma.$$

If we substitute $dc_k/dt$ from the balance equation and use the transformation $\mu\vec{\nabla}J_k/k_BT=\vec{\nabla}(\mu J_k/k_BT)-J_k\vec{\nabla}(\mu/k_BT)$ we get the entropy balance equation with the entropy flux $\vec{j}_S=-\sum_k \vec{J}_k\mu_k/T$; $\mu_k/k_BT=\mu_{0k}/k_BT+(\gamma\Gamma)^2\gamma_k/\gamma c_k(c_k-1)$ and entropy production $\sigma_S=\sum_{k=1}^{r}\{-\vec{J}_k\vec{\nabla}(\mu_k/T)-\mu_km_k/T-\rho\gamma\Gamma^2k_B\gamma_k[Rm_k(dJ_k/dt)/J_k(J_k-Rm_k)-R(dm_k/dt)/(J_k-Rm_k)]\}$. Setting this value $\sigma_S$ equal to $\sigma_S=\sum_{k=1}^{r}J_k^2/L_{kk}+\sum_{r=1}^{s}\sum_{l=1}^{s}R_{rl}\omega_r\omega_l/T$, where $R_{rl}$ are resistivity matrices [3] from the expression $A_r=\sum_{l=1}^{s}R_{rl}\omega_l$, ($s$, number of chemical reactions in a system) we get the general expression relating $dJ_k/dt$, $dm_k/dt$ with $J_k$, $m_k$. Further, considering $m_kR<<J_k$ (almost pure diffusion) one can get an equation for $J_k$, similar to (9) for $q$. Vice versa, when $J_k<<m_kR$ (that is chemical reactions only) one can get an equation for $m_k$ and $\omega_j$. The stability conditions are written similarly to the heat conductivity equation.

## V. STATIONARY NONEQUILIBRIUM STATES

We suppose that the expressions (20), (21) which are results of the present paper are satisfied for arbitrary values of fluxes. Thats why they are believed to be capable of describing the stationary nonequilibrium states far from equilibrium in a more accurate fashion than *EIT* does. Linear deviations should thus be calculated starting from some reference states determined in this case by the stationary values of fluxes. For the distribution (16) one should generate the expansion round about the value

$x_{0x}=x_{st}=q_0R^2t_0/E_0; \quad E_0=E_{st}; \quad q_0=q_{st}; \quad \gamma_0=\gamma_{st}=t_0q_0R^2/\Gamma_0E_0$.

The expression for the entropy thus has a form

$S/k_B = S_\beta/k_B + x_0/(1+x_0) - \ln(1+x_0) - (x-x_0)x_0/(1+x_0)^2 + (x-x_0)^2(x_0-1)/2(1+x_0)^3 + \ldots$.

If one introduces the lifetime in a stationary nonequilibrium state $\Gamma_1(\gamma_0)=\Gamma_0/(1+\gamma_0\Gamma_0); \quad x_1=\gamma_0\Gamma_1(\gamma_0)=x_0/(1+x_0)$, then the latter expression is cast as

$S/k_B = S_\beta/k_B + x_1 + \ln(1-x_1) - (x-x_0)x_1(1-x_1) + (x-x_0)^2(2x_1-1)(1-x_1)^2/2 + \ldots$.

## VI. CONCLUSION

The assumption about the physical systems living for a finite period of time which was the starting point of the exposed here theory allows one to get the mesoscopic theory of the stationary nonequilibrium states at any deviation from the equilibrium. For the method applied it is essential to have the relation of $\Gamma$ as a slave process to the master process $E(t)$. But the concept of the lifetime has a more profound physical sense to cast in one fashion the Newtonian approach to the absolute time and the concepts of time-generating matter. The lifetime parameters is a compromising concept uniting in itself the properties of ordinary dynamical values like energy and the particle number are and the coordinate variables like the time variable is. Mathematically introducing lifetime means yielding an additional information on the stochastic process besides its stationary distribution leaning upon the stationary properties of the slave process.

Introducing the random lifetime as a thermodynamical parameter is analogous to the procedure of introducing an infinite number of fluxes $I^{[r]}$, $r=1,2,...$ [24, 25, 26, 27]. Increasing the number of fluxes corresponds to considering an hierarchy of lifetimes, each truncated description yielding more precise expression for the lifetime. In the suggested scheme of nonequilibrium thermodynamics with finite lifetime those truncated descriptions find their counterpair in the hierarchy of approximations for the thermodynamical conjugate to the lifetime value $\gamma$: to each value of $\gamma^{[r]}$ we set into correspondence the mean lifetime value $<\Gamma^{[r]}>$.

If one integrates (4) and (7) over $\Gamma$, one gets the distribution $P(E)$ depending on $E$ as well as on $\beta$ and $\gamma$ as parameters:

$$P(E)=\int P(E, \Gamma=y)dy=exp\{-\beta E-\gamma \tau_\gamma\}\omega(E)/Z(\beta, \gamma), \qquad (31)$$

where $\gamma \tau_\gamma = -lnL(\gamma,E)$; $L(\gamma, E)=\int e^{-\gamma y}\omega(E,\Gamma=y)dy/\omega(E)$. Averaging the expression for $lnL(\gamma,E)$, we shall get for $\tau = -<lnL(\gamma, E)>/\gamma$; $<lnL(\gamma, E)> = \int exp\{-\beta x - \gamma y\} \omega(x, y) lnL(\gamma, x)dxdy/Z(\beta, \gamma) = \int exp\{-\beta x\}\omega(x)L(\gamma, x) lnL(\gamma,x)dx/Z(\beta,\gamma)$. Let us introduce the value $\tau_0 = <\Gamma> -\tau = \int exp\{-\beta x\}\omega(x) [ L lnL - \gamma \partial L /\partial \gamma ] dx /\gamma Z(\beta, \gamma)$, which vanishes at $\gamma =0$. The value $S – S_\beta = \Delta S = \gamma \tau_0$ ($<0$) is the loss of information when passing from the distribution (7), (4) to (31).

Essential assumption is that the values $\Gamma_0$ and $\ln L(\gamma,E)$ does not depend upon the initial value of $E=x$, that is the value $\Gamma_0$ can be understood as an average on $<E>$ rather than of dynamical random variable $E$. This fact is the direct consequence of the Markoff chain subject to the unperturbed process being equally ergodic and possessing a single (for the distribution (16)) class of the ergodic states (Appendix 2). Then the entropy $S_{<E>}=-<ln\rho(z; E)> = \beta<E> + \gamma \tau + lnZ(\beta, \gamma)$ as an average of the distribution (31) coincides with $S_\beta$ that is the correspondence principle holds: the distribution (31) coincides with the Gibbs one. In the Section 2 (expressions (3-9)) we performed the splitting of the phase space into the cells $(E, \Gamma)$ giving up the equiprobable distribution of points in $\omega(E)$. But although we are now ascribing different weights to different points, collecting them together (summation over $\Gamma$) yields the Gibbs distribution.

The entropy (20) behaves for $\gamma < 0$ and $\gamma > 0$ in a different fashion, this property corresponding to the irreversibility since different signs of $\gamma$ mean different signs of the fluxes $q$. The Gibbs theorem and *H*-theorem [18] hold: the entropy maximum indeed is attained at zero deviations from the equilibrium $\gamma = 0$. The irreversibility in this approach appears as a consequence of the lifetime finiteness hypothesis.

If one compares the exposed thermodynamics with *EIT*, following differences can be outlined:

1. Different expressions for the nonequilibrium temperature, entropy $S$, entropy production (which yield the *EIT* expressions as a particular case for small $\vec{q}$).

2. A new variable of the system size is introduced which should play certain part in the nonequilibrium case. For the continuous description this might be the size of the "continuous medium point" [18]. Real physical systems have finite sizes and finite lifetimes.

3. Explicit expressions for the lifetime $\Gamma$ and its thermodynamic conjugate $\gamma$ are obtained.

## APPENDIX 1. MARKOVIAN STOCHASTIC MODEL

Let the process $y(t)$ be markovian. For the kinetic coefficients

$$K_{\alpha_1...\alpha_m}(y) = \lim_{\tau \to 0}[\tau^{-1} <\Delta y_{\alpha_1}...\Delta y_{\alpha_m}>_y]\ ; \quad \Delta y = y(t+\tau) - y(t)\ ,$$

the potential [23] is written:

$$V(\theta, y) \equiv \sum_{m=1}^{\infty} \beta^{m-1}(1/m!) \sum_{\alpha_1,...\alpha_m} K_{\alpha_1\cdots\alpha_m}(y)\ \theta_{\alpha_1}...\ \theta_{\alpha_m}\ , \qquad (A1.1)$$

where $\beta^{-1}$ is the small parameter; for the equilibrium Gibbs system $\beta = 1/k_B T_{eq}$, $k_B$ being the Boltzmann constant, $T$ is the temperature.

The forward kinetic equation for the distribution density

$$\partial p(y, t)/\partial t = N_{\partial, y} \Phi(-\partial/\partial y, y) p(y, t)\ , \qquad (A1.2)$$

where $\Phi$ is the stochastic potential, $\Phi(\theta, y) = \beta V(\theta/\beta, y)$, the operator $N_{\partial,y}$ defines the order of operations (differentiating on $y$ goes the last).

The Laplace transform gives the equation for $Q$ (10):

$$\partial Q(\exp\{-\theta\}, t)/\partial t = N_{\theta, \partial/\partial\theta} \Phi(-\theta, -\partial/\partial\theta, t) Q(\exp\{-\theta\}, t)) . \qquad (A1.3)$$

One can show that the lifetime (2) is governed by the Hermitian conjugate operator and for $L$ (11) following holds:

$$N_{x,\partial} \Phi(\partial/\partial x, x) L(x, s) = s L(x, s) \qquad (A1.4)$$

with the condition $L(0, s) = L(x, 0) = 1$. For the nonhomogeneous Markoff systems if $L$ depends on the initial time $t_0$ the equation (A1.4) is generalized:

$$s L_{t_0}(x, s) = \partial L_{t_0}(x, s)/\partial t_0 + s N_{x,\partial} \Phi(\partial/\partial x, x, t \to t_0 - \partial/\partial s)([L_{t_0}(x, s) - 1]/s) . \qquad (A1.5)$$

From these expressions the precise solutions for the Markoff stochastic models of the physical systems can be found. We note bypassing that in (A1.1-A1.5) the coefficients are unperturbed equilibrium values dependent on $\beta_0$, $\gamma_0$, but not on $\beta$, $\gamma$.

**APPENDIX 2. ALGORHITHMS OF THE PHASE COARSENING OF THE COMPLEX SYSTEMS**

One can represent an open thermodynamic system as evolving in a random medium whose mathematical model will be either Markoff renewal processes or semimarkoff processes. The local characteristics of the system depend on the random medium state. The scheme of the asymptotic phase coarsening present the simplified description of the system evolution in a random medium which can be performed basing upon a simple set of heuristic rules [19]. The semimarkovian process to describe the evolution of a random system is considered in the enlarged time scale.

In our case the stationary distributions are described by the Gibbs distributions. Absorbing state is the degenerated state of the system with $E=0$, and the degeneracy probability is $P_0=1/Z(\beta)$. The ergodic Markoff chain has the distribution $\rho$. The residence times in the states $x$, $\theta_x$, are given by the distribution functions $G_x(t)=P\{\theta_x \leq t\}=P\{\theta_{n+1}\leq t|\xi_n=x\}$ ($\xi_n$ are the states of the Markoff chain) and by the average residence times $m(x)=\int_0^\infty (1-G_x(t))dt=E\theta_x$, which are limited: $0 \leq m(x) < \infty$. Lets assume that a real system has besides the class of the ergodic states the class of the trapping states where the absorbing of the Markoff chain is possible. The phase space of a real system is then $X=X^0 \cup X' \cup X_0$, where $X^0$ is the class of the ergodic states of the reference enclosed Markoff chain (without taking into account the absorption into $X_0$); $X_0$ are trapping states of an enclosed Markoff chain, $X'$ is the finite set of the states. In [19] it was shown that the coarsened random process is Markovian with two states $X_{0coars}$ and $X^0_{coars}$. The residence time in the stable state is distributed exponentially with the parameter

$$\lambda_1=1/m; \quad m= \int_{X^0} \rho(dx)m(x)/P_{10}; \quad P_{10}=P\{\xi_{n+1, coars}=X_{0 coars}|\xi_{n, coars}=X^0_{coars}\} \quad .(A2.1)$$

Thus the expression (16) is obtained in which $m=\Gamma_0$. The Markoff processes as objects modelling a complex system appear not as a *a priori* hypothesis but rather as a result of splitting phase space of states and "gluing up" together the states which belong to one and the same class. The algorhitm of the phase coarsening shows the natural property of the complex systems: the transitions in a complex system between the classes of states are rather governed by the Markoff property. If one neglects the detailed description of the system evolution (neglecting the transitions within a class) then (at the condition of big enough residence time within a class) the system losses the dependence of the inter-class transitions on the behaviour inside of a class, and the residence time in a class is set as a sum of a big (random) number of the random values which are random times of residence in the states; under certain conditions this residence time in the class can be considered to have an exponential distribution. We have:

$$P(\varGamma_G > t) = exp\{-t/m\}; \quad P(\varGamma_x > t) \approx exp\{-t/m\}, \qquad (A2.2)$$

where $\varGamma_G$ is the residence time of a coarsened system before absorption; $\varGamma_x$ is the residence time within the class $X^0$ with the initial state *x*. The error in the approximation (A2.2) is proportional to the degeneracy probability of the Gibbs system $P_0 = 1/Z_\beta$.